# BEAM SCIENCE
## AND
# TECHNOLOGY

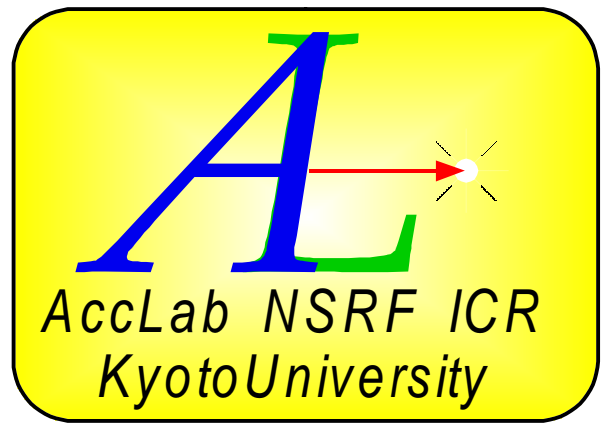

**STUDY OF MULTIPLE-BEAM RFQ**

V. Kapin, M. Inoue, Y. Iwashita and A. Noda,
Accelerator Lab., Inst. Chem. Res., Kyoto University,
Gokanosho, Uji, Kyoto 611, Japan

# STUDY OF MULTIPLE-BEAM RFQ

V. Kapin, M. Inoue, Y. Iwashita and A. Noda,
Accelerator Lab., Inst. Chem. Res., Kyoto University,
Gokanosho, Uji, Kyoto 611, Japan

*Abstract*

There are applications, which require MeV-range multiple-beams consisting of a large number of identical highly packed beamlets. The multiple-beam RFQ (MB-RFQ) arranged as a matrix array of longitudinal rod-electrodes is appropriate candidate. A configuration of MB-RFQ resonator should ensure identical quadrupole fields in every accelerating channel.

The MB-RFQ resonators based on TEM transmission lines are studied. The resonators are designed by a periodical multiplication of a 4-rod unit cell in transverse direction. To understand fields of resonator the normal mode technique is applied. Novel configurations of MB-RFQ resonators based on these normal modes are generated. The RF properties of resonators are verified with computer simulations done with MAFIA code.

Beam dynamics study for the initial 400-keV MB-RFQ has been performed using PARMTEQ code. The MB-RFQ parameters and the results of beam dynamics simulations are presented. The calculated beam transmission is 33% at injection of 50-keV, 20mA deuteron beam.

## I. ITRODUCTION

In recent years, many ion sources with a broad-beams have been developed [1-5]. The broad-beams are formed as multiple-beams consisting of an array of identical single-beams (or beamlets). The beamlets are packed very closely. The packing factor defined as a ratio between the sum of areas occupied by beamlets and a total area of broad-beam can reach 40-50 %. The number of beamlets may achieve several hundreds (or even thousands). Transverse size of broad-beam can be up to one meter or more.

There are applications, which require MeV-range broad-beams, e.g. heating of plasmas in magnetic confinement devices [3-5]. The typical required ion flux is 20-30 mA/cm$^2$ over a total surface of about 200cm$^2$. Existing broad-beam accelerators use the electrostatic method of acceleration. Because of technical problems due to the voltage breakdown, an attainable level of beam energy is restricted. Up to now, a D$^-$ current of 1A (20mA/cm$^2$) 50 keV energy was obtained [5].

Perhaps, some RF-acceleration method adapted to a multiple-beam acceleration may be applied for MeV-range broad-beam accelerators. Many RF multiple-beam accelerating structures had been already presented in the past years [6-15].

MB-RFQ arranged as an array of longitudinal rod-electrodes (see Fig. 1) is known for a long time [6,9,11]. It is required to realize a high packing factor without disturbing RFQ fields and excite correctly electrodes surrounding every channel. In this paper, we try to find out possible configurations of the MB-RFQ structures satisfying the above requirements.

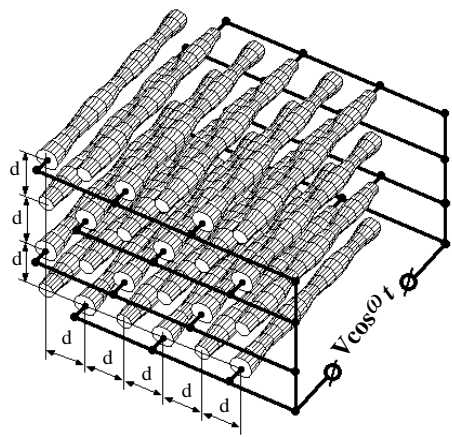

Figure 1: The matrix-array of RFQ electrodes.

MB-RFQ structures are treated as the TEM transmission line resonator. The resonators are designed by a periodical multiplication of a 4-rod unit cell in transverse direction. The normal mode technique is applied to decompose a complicated field resonator into fields of normal modes.

The resonator field is expanded into normal modes having simple field patterns. In general case of resonator with an arbitrary number of rods and normal modes, analysis becomes difficult. To simplify problem, only normal modes composed from four normal modes of a 4-rod unit cell are selected. Similar to normal modes of a 4-rod unit cell (coaxial, quadrupole and two dipole modes), selected normal modes of multi-rod resonator have clear field patterns.

## II. NORMAL MODES OF THE NCSTL.

Let us consider MB-RFQ resonator with *N* longitudinal electrodes as a resonator based on the *N*-conductor shielded TEM transmission lines (NCSTL). The propagation of TEM waves is described by the system of the telegraph equations. To facilitate solution of the telegraph equation the normal mode technique is usually applied. The resonator field is expanded into normal modes having simple field patterns. In the transmission line with *N*-conductors, there are *N* normal TEM modes.

13

This technique has been applied to study 4-rod RFQ using the four-conductor shielded transmission line (4CSTL) [16-18]. In the case of $N > 4$, the analytical definition of normal modes becomes difficult. To facilitate the study, let us restrict a number of considered normal modes in the NCSTL. A 4-rod configuration with known normal modes is considered as a unit cell. Normal modes of the NCSTL are composed by a periodical multiplication of a 4-rod unit cell in transverse direction.

The normal modes in the 4CSTL can be observed in the resonator shown in Fig.2. All four electrodes are grounded at the same longitudinal position, $z=0$ and have open ends at the another end of resonator, $z=l$.

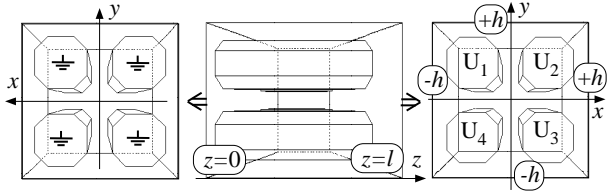

Figure 2: 4-rod RFQ allowing an observation of normal modes.

Figure 3 shows the E-line patterns of TEM normal modes in the 4CSTL calculated with MAFIA code [19]. Three different combinations of two boundary conditions (perfect conductor, $E_t = 0$ or infinitely permeable, $H_t = 0$) on the shield are presented. The first and second type has four normal modes (coaxial, quadrupole and two dipole). The third type with $H_t = 0$ on all four sides of the shield has only three modes, because it corresponds to open transmission line.

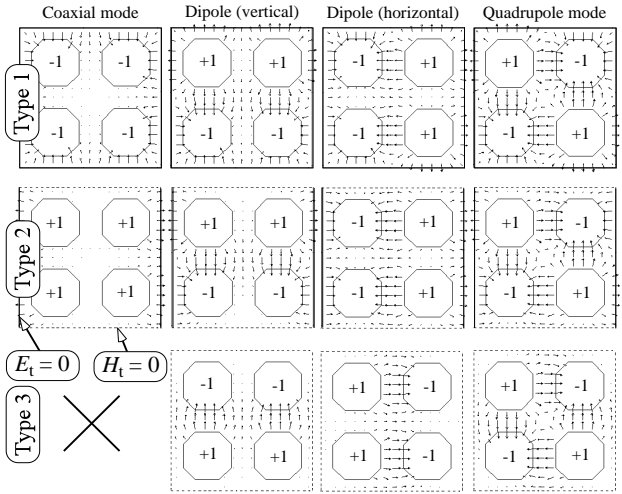

Figure 3: The E-line patterns of normal modes in the 4CSTL for three types of boundary conditions on shield.

The normal modes in the NCSTL can be observed by the similar way. The MB resonator with 4x4-matrix array of electrodes is shown in Fig.4. All electrodes are grounded at the same longitudinal position, $z=0$ and have open ends at the another end of resonator, $z=l$.

Figure 5 shows the E-line patterns of normal TEM modes calculated by MAFIA code. Three types of boundary conditions on the shield are presented. The amplitude values of conductor potentials are shown on the conductor cross-sections.

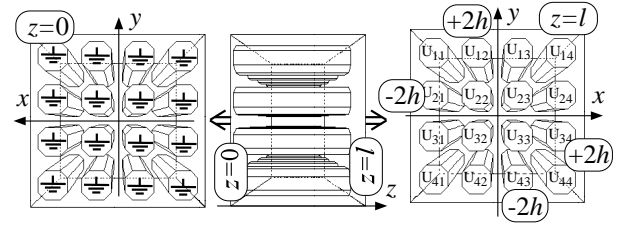

Figure 4: MB-RFQ resonator with a 4x4-matrix array of electrodes allowing an observation of normal modes.

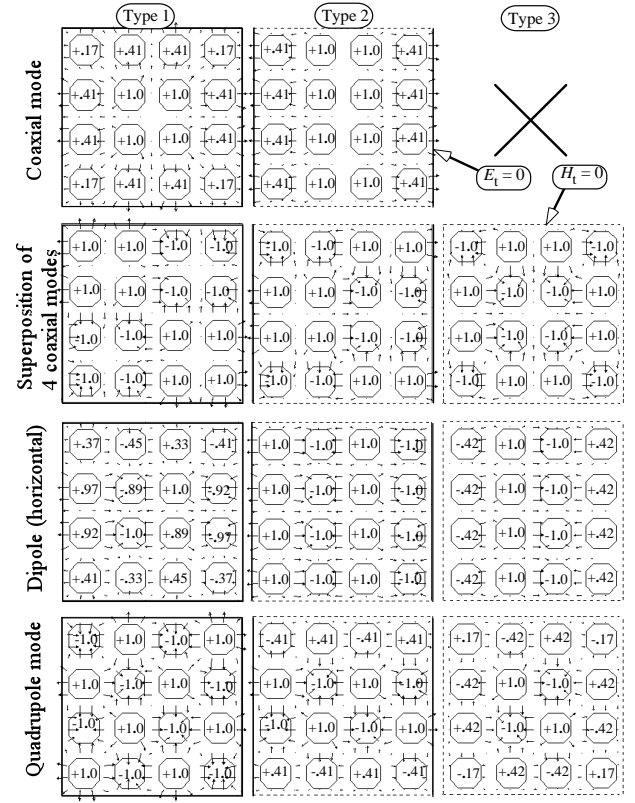

Figure 5: The E-line patterns of normal modes in the MB-RFQ resonator for three types of boundary conditions on the shield.

## III. SINGLE-MODULE RESONATORS BASED ON A CONVENTIONAL RFQ.

The quadrupole mode of the type 1 has correct-balanced quadrupole potentials for all channels. The field of the quadrupole mode increases sinusoidally along z-direction. Being excited on this quadrupole mode, the MB-resonator shown in Fig. 4 can be used as an initial matching section.

Figure 6 shows the MB resonator designed using an extension of 4-rod RFQ in the transverse direction. All electrodes are divided into two groups in a chess order. The electrodes of two groups are grounded in opposite manner. Similar structures based on this design principle had been considered in several works by other authors [6,9,10,12].

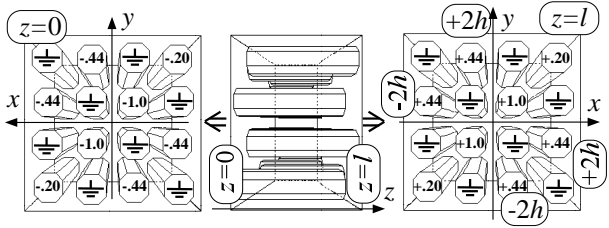

Figure 6: MB-RFQ resonator composing by an extension of 4-rod RFQ.

Figure 6 also shows voltage amplitudes at the middle of the resonator, $z=l/2$. The voltages on the electrodes surrounding RFQ-channels deviate from quadrupole symmetry. To explain a reason of such resonator behavior, the field in this MB-resonator can be interpreted in terms of normal mode technique. The field of the original 4-rod resonator is described by combination of quadrupole and coaxial mode, which has been presented in [17,18]. In contrast to 4-rod resonator, the coaxial mode of MB-RFQ (see Fig.5, Coaxial mode of type 1) has unequal potentials of electrodes. Therefore, the combination of qurupole with this coaxial mode in the MB resonator does not provide correct voltages on the electrodes surrounding RFQ-channels. It is difficult to adapt this type of resonator to MB-RFQ-acceleration.

The well-balanced fields can be provided in the different MB resonator. It is shown in Fig. 7. This MB resonator designed using a periodical multiplication of the 4-rod unit cell in the transverse direction. The field of the 4-rod unit cell of the resonator is described by combination of the quadrupole mode and the normal mode shown in the second row of the first column of Fig.5. Since composing modes have well-balanced amplitudes, the total voltages on electrodes are balanced well and every second RFQ-channel can be used for an acceleration of beamlets.

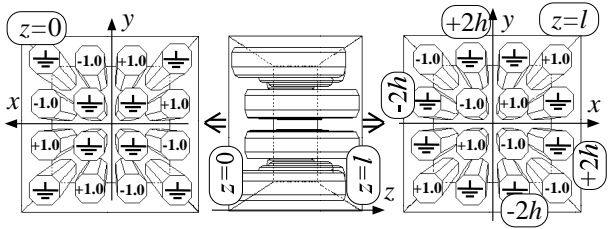

Figure 7: MB-RFQ resonator composing by a periodical multiplication of 4-rod unit cell.

## IV. PROBLEMS OF MULTI-MODULE RESONATORS

Considered in the previous section MB-RFQ structures consist of a single module. In a single-module resonator, one end of electrodes is either connected to or cut at a tank bottom. To realize a multi-module structure, electrodes should continuously spread along of total tank. The support of electrodes is provided by transverse stems. It was also proposed in works [9,12] to use a coupling elements to equalize potentials of electrodes.

In conventional RFQ structure, adjacent electrodes have different potentials along the whole structure. They can not be interconnected by short-circuited couplers. Only next adjacent electrodes may be interconnected "in a chess order". This circumference affects on a maximum value of the packing factor, which can be achieved in a realistic design.

Two possible configurations of couplers consisting of ideal thin conductors are shown in Fig. 8. In the first (left) configuration, coupling conductors bypass the RFQ-channels. Only a quarter of RFQ-channels is used. The second (right) configuration allows the penetration of the coupling elements inside RFQ-channels. The couplers have aperture-holes. The original RFQ-fields are distorted in this configuration.

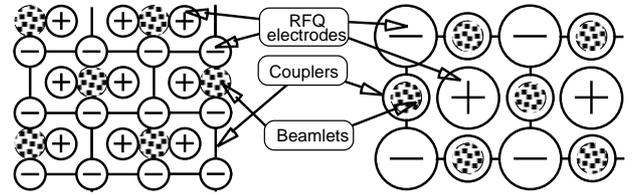

Figure 8: Two possible configurations of coupling elements.

In order to prevent voltage breakdown, the realistic configurations should be composed of thick conductors. The curvature radii of all conductors could not be less than the curvature of RFQ-electrodes. We tried to design such couplers. However, the packing factor is reduced to 5% and 10% for the left and right configurations, respectively. These low values of the packing factor do not suit to values of existing MB ion sources.

## V. MB-RFQ BASED ON A NEW 4-ROD RFQ

Let us find out a type of a resonator, which allows a short-circuited connections between adjacent electrodes. It may to be helpful to avoid a reduction of the packing factor in multi-module structures with transverse stems.

Conventional 4-rod RFQ structures based on quadrupole mode only or on a combination of quadrupole and coaxial modes, and dipole modes are usually considered as unwanted distortions, which must be suppressed. Let us consider an unusual combination of normal modes including a dipole mode [20]. This is a superposition of quadrupole mode and dipole mode, which are summed at a $\lambda/4$ phase shift between each other. Figure 9 shows the voltage distributions for this combination of quadrupole and dipole modes in the case of the 4-rod unit cell. The result of the summation is given on the right side of the Fig. 9.

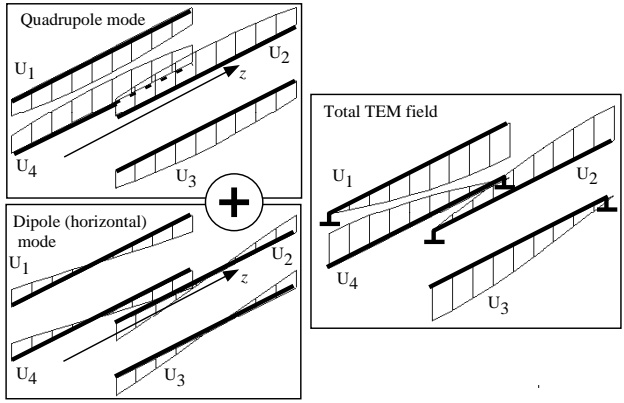

Figure 9: The superposition of quadrupole and dipole modes shifted by $\lambda/4$ to each other.

A unit module of such resonator is shown in Fig. 10. An example of the multiple-module MB-RFQ resonator with 6x6-matrix array of electrodes is shown in Fig.11.

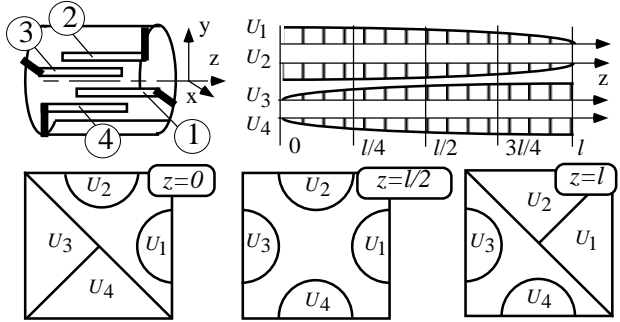

Figure 10: Unit module of a new 4-rod resonator, voltage distributions on its conductors, and pole tips at different cross-sections of the resonator.

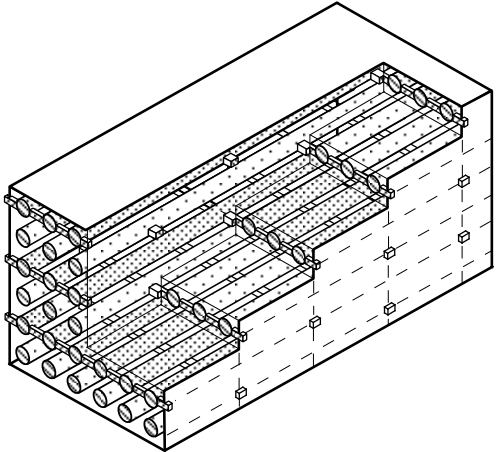

Figure 11: MB-RFQ resonator with 6x6 matrix array of the electrodes.

To provide RFQ-acceleration, the field at the resonator axis should be the same as in a conventional RFQ-channel. It is described by the following lowest-order electric-field potential function [21,22]:

$$U(r,\psi,z) = (V/2) \cdot [\kappa (r/a)^2 \cos 2\psi - AI_0(kr)\sin kz], \quad (1)$$

where

$$A = \frac{(m^2-1)}{m^2 I_0(ka) + I_0(mka)}, \quad \kappa = 1 - AI_0(ka), \quad k = \frac{2\pi}{\beta\lambda}.$$

At every cross-section of the resonator, electrodes have some definite voltages $U_1$, $U_2$, $U_3$, and $U_4$. To provide the RFQ fields described by Eqs. (1), surfaces of the RFQ pole tips for every electrode are defined by equation:

$$U(r,\psi,z) = U_i, \quad i = 1,...4. \quad (2)$$

Figure 10 shows pole tips at the ends and the middle of the resonator. The length of the unit module of the resonator, $l$ is about a quarter of the wavelength, $\lambda$. Apart from conventional modulation of RFQ electrodes with period length, $\lambda_{RFQ}=2\beta\lambda$, the pole tips are modulated with the period length, $\Lambda_{RFQ}=2l$. The RFQ-channel has a zero optical transparency, and can not provide a conventional RFQ acceleration on z-axis. The beam dynamics must be modified in this case.

We have revealed a possible way to provide RFQ acceleration in this accelerating channel [20]. It is based on the fact that parts of beam coherently oscillate around the axis during transverse oscillations in a strong-focusing quadrupole channel. Figure 12 shows particle motion in transverse plane, $X0Y$ calculated by PARMTEQ code [22] for a conventional RFQ. Part of beam injected in the first quadrant of the $X0Y$-plane performs periodical excursions between the first and the third quadrants of $X0Y$-plane. This part of the beam can be transported in the accelerating channel of the new resonator, if the space period of transverse beam oscillations is matched to period of electrode modulation $\Lambda_{RFQ}$.

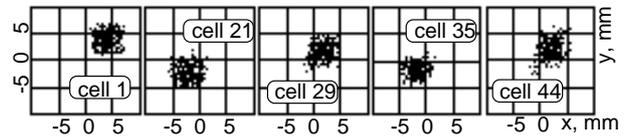

Figure 12: Beam structure in the transverse plane $X0Y$ at different cells of a conventional RFQ-channel.

In accelerating channel of the new resonator, the beam with the velocity, $v=\beta c$ should be injected with a shift from the RFQ axis. The beam will oscillate coherently around the quadrupole axis with the frequency of betatron oscillations, $\Omega_R$. When the wavelength of the betatron oscillations, $\Lambda_R=2\pi\beta c\Omega_R$ is matched to the period length of the pole tips modulation, i.e. $\Lambda_R=\Lambda_{RFQ}/(2n+1)$, the beam bends round the pole tips performing a kind of a "slalom" motion. Under this condition, the smooth frequency of betatron oscillation, $\mu_R$ is proportional to the relative velocity, $\beta$

$$\mu_R \cong 8\pi(n+1/2)\beta \quad (3)$$

The last equation defines the increasing dependence of the radial focusing forces along the RFQ channel at a fixed number, $n$. To test this principle, numerical simulations have been done. The results of beam dynamics simulations will be presented in the section VII.

16

## VI. RF Tuning of a new MB-RFQ [23].

The RF properties of resonators have been verified with computer simulations done with MAFIA code. Fig.13 shows the MB-RFQ resonator used for MAFIA calculations.

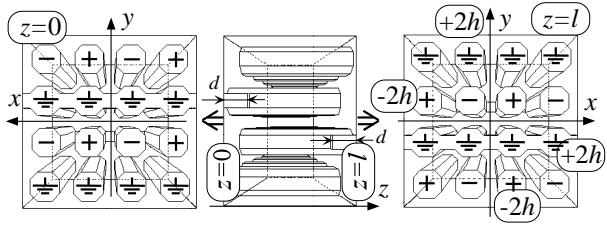

Figure 13: MB-RFQ resonator with 4x4-matrix array of electrodes.

For the case of MB-RFQ, the mode combination shown for 4-rod unit cell (Fig.9) is extended to the superposition of quadrupole mode of type 1 and dipole mode of type 2 (see Fig.5). For a real configuration of MB-RFQ resonator, the boundary conditions on all tank walls must correspond to a perfect conductor. The dipole mode of type 2 is replaced by the dipole mode of type 1. In contrast to the dipole mode of type 2, potentials of electrodes surrounding RFQ-channels for the dipole mode of type 1 deviate from a correct dipole field (see Fig.5). As the result, a total field in the MB-resonator is distorted.

A difference between dipole modes of type 1 and type 2 is appeared as different boundary conditions on the shield for the upper and lower rows of 4-rod unit cells. Free frequencies of the outer cells deviate from their values for cells in the middle rows of MB-resonator. The free frequencies of the outer cells should be tuned.

The free frequency of a 4-rod unit cell depends on an electrical length of conductors. The tuning method has been found. The length of conductors in the middle rows should be less than the length in the outer cells on some value, $d$ (see Fig.13).

Fig. 14 shows the distributions of the quadrupole $V_{i,j}$ and dipole voltages (vertical, $Dv_{i,j}$ and horizontal, $Dh_{i,j}$) in separate $i,j$-channels calculated with MAFIA code. The quadrupole voltage $V_{i,j}$ in the $i,j$-channel is calculated from the voltages of electrodes surrounding the channel by the relation

$$V_{i,j} = (U_{i,j} - U_{i,j+1} + U_{i+1,j+1} - U_{i+1,j})/2$$

The voltages of horizontal ($h$) and vertical ($v$) dipole modes, $Dh_{i,j}$ and $Dv_{i,j}$ are defined by the following formulas:

$$Dh_{i,j} = (U_{i,j} - U_{i,j+1} - U_{i+1,j+1} + U_{i+1,j})/2$$
$$Dv_{i,j} = (U_{i,j} + U_{i,j+1} - U_{i+1,j+1} - U_{i+1,j})/2$$

For the case of $d=0$, the values of quadupole voltage, $V_{i,j}$ are different for the middle and outer channels. The phase shift between quadrupole and dipole voltages deviates from $\lambda/4$ for outer channels.

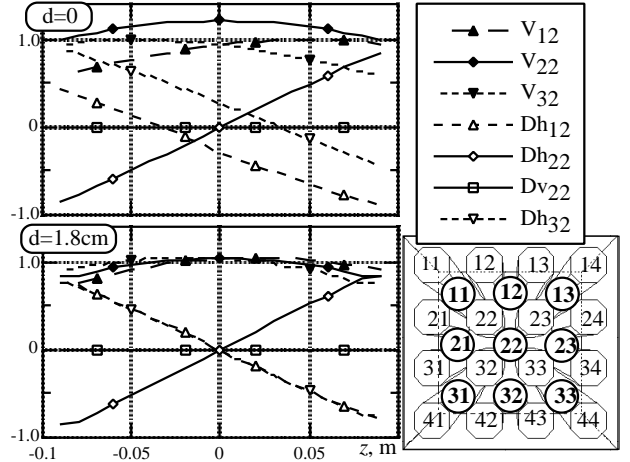

Figure 14: The distributions of the quadrupole and dipole voltages in separate channels calculated with MAFIA code.

The optimal value of $d=1.8$cm has been determined. In the optimal case, curves of quadrupole voltages became very similar and the required phase shift of $\lambda/4$ is restored. Thus, RF fields in different channels of MB-RFQ structure can be equalized to provide identical RF fields in every RFQ-channel.

## VII. Study Beam dynamics in MB-RFQ [24].

### 7.1 400-keV MB-RFQ

In order to evaluate an achievable beam transmission and current for the MB-RFQ, beam dynamics simulations has been performed for a single channel of MB-RFQ linac, which accelerates ions of deuterons from 50 keV to 400 keV at RF frequency 108 MHz.

The 400-keV MB-RFQ linac consists of two different RFQ structures. The first MB-RFQ is a structure with a conventional RFQ acceleration and focusing in a quarter-wave resonator with an increasing RFQ voltage. The second structure is a two-module MB-RFQ resonator for a "slalom"-beam.

The configurations of these structures for 4x4-matrix arrays of electrodes were shown in Fig.4 and Fig.13, respectively. The Fig. 15 shows the cross-sections of a one channel and the voltage distributions on the electrodes of the RFQ resonators.

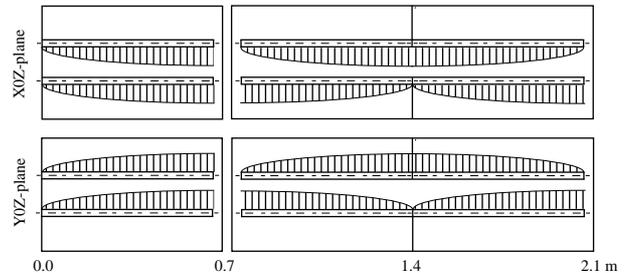

Figure 15: The cross-sections of two RFQ resonators and the voltage distributions on the electrodes.

## 7.2 Numerical code for simulation beam dynamics.

As it has been discussed in the section 5, the acceleration in the MB "slalom"- beam structure has several basic features in comparison with a conventional RFQ:

a double-periodical modulation of RFQ electrodes with period lengths, $\lambda_{RFQ}$, and $\Lambda_{RFQ}$;

the period of the betatron oscillations must be matched to the period length $\Lambda_{RFQ}$;

the modulation coefficient, $m$ must be periodically put be equal to zero with the period $l$ along the RFQ channel;

the voltage distributions are essentially non-uniform along RFQ-channel.

Due to these differences a conventional methods and codes for design RFQ channels like PARMTEQ-code [22] could not be used directly. For calculation of beam dynamics in the MB-RFQ structures, a complementary numerical code has been written using the Digital Visual FORTRAN by Digital Co.

This code works interactively and allows introduce a desirable dependencies of RFQ parameters (voltage distribution $V(z)$, the focusing strength $B$, synchronous phase $\varphi_S(z)$, and ratio between defocusing and focusing force) as smooth spline-functions of the longitudinal coordinate, $z$. Using these functions, the following parameters of the $i_{cell}$-th accelerating cell are numerically calculated: cell length $\lambda_{RFQ}(i_{cell})$, virtual aperture $a_{RFQ}(i_{cell})$, synchronous phase $\varphi_S(i_{cell})$, modulation coefficient $m(i_{cell})$, virtual inter-electrode voltage $V_{RFQ}(i_{cell})$, real apertures for four transverse semi-axes $a_{x>0}(i_{cell})$, $a_{x<0}(i_{cell})$, $a_{y>0}(i_{cell})$, and $a_{y<0}(i_{cell})$.

To analyze beam dynamics in generated RFQ channel, a PARMTEQ code have been modified and adapted. The previously calculated real apertures $a_{x>0}(i_{cell})$, $a_{x<0}(i_{cell})$, $a_{y>0}(i_{cell})$, and $a_{y<0}(i_{cell})$ have been introduced into PARMTEQ code and particles which hit the pole tips of modified RFQ electrodes has been excluded from further beam dynamics simulation.

## 7.3 Results of beam dynamics simulations.

The calculated parameters of MB-RFQ are presented in Fig. 16. The first resonator performs a radial matching, bunching and pre-acceleration of beam from 50keV to 114 keV. The length of the resonator, $l$ is equal to about a quarter of the wavelength $\lambda$, i.e. $l$=0.7 m at the frequency $f_0$=108 MHz. The "slalom"-beam RFQ accelerates a bunched beam at an almost constant synchronous phase up to 400 keV. The "slalom"-beam RFQ section was designed using the condition (3).

The first section has been designed to accept a continuous monochromatic beam with the transverse emittance $\varepsilon = 25\pi$ cm mrad 50keV. The beam is bunched within a short length of structure ($l = \lambda/4$) with a fast increase of the synchronous phase (see Fig.16). For initial K-V beam ($\alpha_{x,y} = 0.8$, $\beta_{x,y} = 15.0$ cm/mrad), the beam transmission is equal to 78% for a zero-current beam. It reduces to 68% and 62% at the beam current 15 mA and 30 mA, respectively. Figure 17 shows the example of the beam-structure and the electrode profiles in the XOZ-plane.

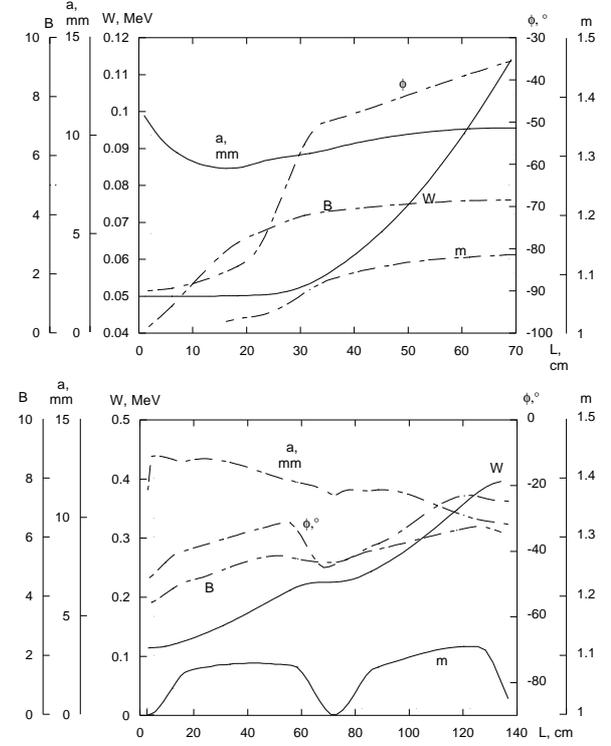

Figure 16: Parameters for the first (the top graph) and the second (the bottom graph) 108-MHz MB-RFQ channels.

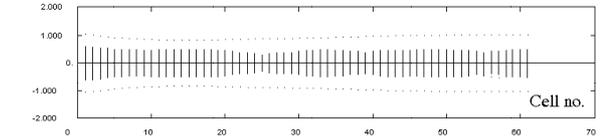

Figure 17: The XOZ cross-section of the RFQ-channel showing the beam structure and the electrode profiles of the first section.

The simulation of the "slalom"-beam RFQ has been made for beam with the energy spread $\pm 5$ keV, the phase spread $\pm 50°$, and the transverse emittance $\varepsilon = 15\pi$ cm mrad. For injected K-V beam ($\alpha_{x,y} = \mp 0.8$, $\beta_{x,y} = 8.0$ cm/mrad) the beam transmission is equal to 66% for a zero-current beam. At the beam current 15 mA and 30 mA the beam transmissions become to 63% and 53%, respectively. Figure 18 shows the example of the beam motion and the electrode profiles in the XOZ-plane. Similar situation occurs for XOY-plane. Beam performs "slalom" motions, avoiding pole tips.

Finally, common transmission of two structures has been calculated. The transmission of 42% for a zero-current beam reduces to 33% at the beam current of 20 mA, respectively. Figure 19 shows the phase spaces of the output beam for the latter case.

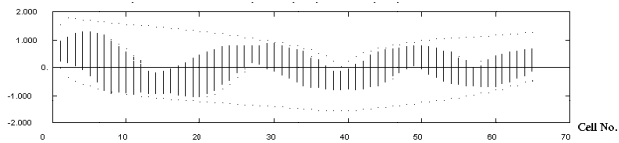

Figure 18: The *X0Z* cross-section of the RFQ-channel showing the beam structure and the electrode profiles.

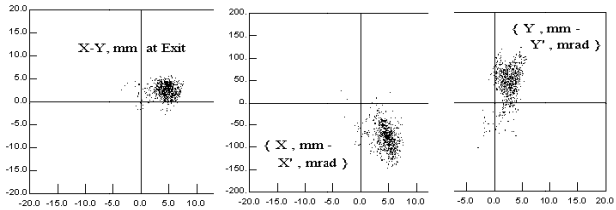

Figure 19: The phase spaces of the 400-keV beam at the injection current of 20 mA.

## ACKNOWLEDGEMENTS.

This study has been conducted under the Postdoctoral Fellowship Program for Foreign Researchers of Japanese Science Promotion Society and supported in part by the Grant-in-Aid for JSPS Fellows from the Ministry of Education, Science, Sports and Culture of Japan.

## VIII. REFERENCES

[1] A.Forrester, Large ion beams: fundamentals of generation and propagation, 1988 by John Wiley & Sons, Inc.

[2] I.G.Brown, Rev.Sci.Instrum., 65(10), 1994, p.3061.

[3] Y.Okumura et al., Rev.Sci.Instrum., 67(3), 1996, papers on p.1018, 1092.

[4] Proc. of the 7th Int. Conf. On Ion Sources, Rev.Sci.Instrum. Vol.69(2), 1998, papers pp. 843, 846, 863, 874, 880, 899, 908, 920, 929, 947, 956, 977, 986.

[5] Simonin, Proc. of the Sixth European Particle Accelerator Conference (EPAC'98), 1998 by IOP Publishing Ltd., pp.293, 699.

[6] D.A.Swenson, Proc. of the Heavy Ion Fusion Workshop, Berkeley, 1979, LBL-10301, SLAC-PUB-2575, p. 239.

[7] Ion Linear Accelerators, ed. by B.P.Murin, Vol. 1, Atomizdat, Moscow, 1978 (*in Russian*).

[8] R.M.Mobley et al, IEEE Trans. on Nucl. Sci., Vol. NS-28, No.2, 1981, p.1500.

[9] E.V.Gromov et al., Inventor's Certificate of USSR No.831044, Bulletin of inventions, 1985, No. 12 (*in Russian*).

[10] H.Klein et al., Proc. of the 1981 Linear Accelerator Conference, Santa Fe, LANL report LA-9234-C, p.96.

[11] P.Junior et al., IEEE Trans. on Nuclear Science, Vol. NS-30,No. 4, 1983, p.2639.

[12] E.V.Gromov and S.S.Stepanov, Proc. of MEPhI's collected papers, Moscow, Energoatomizdat, 1983, p.35 (*in Russian*).

[13] R.J.Burke et al., IEEE Trans. on Nuclear Science, Vol. NS-32, No. 5, 1985, p.3347.

[14] R.J.Burke et al, Nucl.Instrum&Meth., Vol. B10/11 (1985) p.483.

[15] J.Madlung et al., Proc. 1995 Particle Accelerator Conf., Dallas, 1996 by IEEE, p.908.

[16] V.Kapin, Proc. of the Fourth European Particle Accelerator Conference, 1994 by World Scientific Publishing Co. Pte. Ltd., Vol.3, pp.2191.

[17] V.Kapin, M.Inoue, Y.Iwashita and A.Noda, Proc. of the 1994 Int. Linac Conf., 1994 by KEK, Japan, Vol.1, pp.254.

[18] V.Kapin, M.Inoue, Y.Iwashita and A.Noda, Bull. Inst. Chem. Res., Kyoto Univ., Vol.73, No.1, 1995, pp.50.

[19] R.Klatt et al., Proc.1986 Linear Accelerator Conf., 1986 by SLAC, p. 276.

[20] V. Kapin, A. Noda, Y. Iwashita and M. Inoue, Proc. of XVIII International Linear Accelerator Conf., CERN, Geneva, 1996, p.722.

[21] I.M.Kapchinskiy, Theory of resonance linear accelerators, 1985 by OPA (Amsterdam).

[22] K.R.Crandall et al., Proc. of the 1979 Linear Accelerator Conf., 1980 by Brookhaven Nat. Lab., BNL-51143, p.205.

[23] V. Kapin, M. Inoue, Y. Iwashita and A. Noda, to be published in Proc. of XIX International Linear Accelerator Conference, Chicago, Illinois, USA, August 23-28, 1998.

[24] V. Kapin, M. Inoue, and A. Noda, Sixth European Particle Accelerator Conference (EPAC'98), 1998 by IOP Publishing Ltd., p.725.